\documentclass{elsart}
\usepackage{natbib}
\usepackage{amssymb}

\input{epsf.sty}						

\newenvironment{figMacPc}[4]				
{\begin{figure}[h]						
	\epsfxsize=#2\centerline{\epsfbox{#1}}		
	\caption{#3}							
	\label{#4}							
}
{\end{figure}}
\newenvironment{tabMacPc}[3]				
{\begin{table}[htb]							
	\caption{#1}							
	\begin{center}
	{#2}									
	\end{center}
	\label{#3}							
}
{\end{table}}

\newcommand{\numdim}[2]{$#1\,(\mbox{#2})$}
\newcommand{\conc}[1]{\,[\mbox{#1}]}
\newcommand{\indice}[1]{${#1}=1,2,3$}
\newcommand{\indici}[2]{${#1,#2}=1,2,3$}

\newtheorem{proposizione}{Proposition}[section]
\newtheorem{corollario}{Corollary}[section]
\newcommand{\cvd}						
{\par\raggedleft{\rule{2mm}{2mm}}}		%
\newenvironment{dimos}					
{\begin{trivlist}\item[\hspace{\labelsep}	
{\sc\noindent Proof.}]}					
{\cvd\end{trivlist}}					
\newcommand{\bdim}{\begin{dimos}}	 		
\newcommand{\edim}{\end{dimos}} 	 		
\begin{document}
\begin{frontmatter}
\title
{
A chemically driven fluctuating ratchet model for actomyosin interaction\thanksref{label:title}}
\thanks[label:title]{This work has been performed within a joint cooperation agreement between Japan Science and
Technology Corporation (JST) and Universit\`a di Napoli Federico II, under partial support by
Gruppo Nazionale di Calcolo Scientifico (GNCS) of Istituto Nazionale di Alta Matematica (INdAM).}
\author[Osaka]{T. Shimokawa\corauthref{cor}},
\corauth[cor]{Corresponding author. Tel: +81-6-6850-6536, Fax: +81-6-6850-6557}
\ead{simokawa@bpe.es.osaka-u.ac.jp}
\author[Osaka]{S. Sato},
\ead{sato@bpe.es.osaka-u.ac.jp}
\author[Italy]{A. Buonocore},
\ead{aniello.buonocore@unina.it}
\author[Italy]{L.M. Ricciardi}
\ead{luigi.ricciardi@unina.it}
\address[Osaka]{Graduate School of Engineering Science, Osaka University,\\
1-3 Machikaneyama Toyonaka Osaka, JAPAN}
\address[Italy]{Dipartimento di Matematica e Applicazioni, Universit\`{a} di Napoli Federico II,\\
Via Cintia 80126 Napoli, ITALY}
\begin{abstract}
With reference to the experimental observations by T. Yanagida and his co-workers on actomyosin interaction, a Brownian motor of fluctuating ratchet kind is designed with the aim to describe the interaction between a Myosin II head and a neighboring actin filament. Our motor combines the dynamics of the myosin head with a chemical external system related to the ATP cycle, whose role is to provide the energy supply necessary to bias the motion. Analytical expressions for the duration of the ATP cycle, for the Gibbs free energy and for the net displacement of the myosin head are obtained. Finally, by exploiting a method due to Sekimoto (1997, J. Phys. Soc. Jpn., 66, 1234), a formula is worked out for the amount of energy consumed during the ATP cycle.
\end{abstract}

\end{frontmatter}
\section{Introduction}
The sliding process of a myosin head along the actin filament has been experimentally found by T. Yanagida and his co-workers to occur according to a random number of steps of constant size, mainly in a preferred direction. This observation  is
undoubtedly in strong support of the loose-coupling relation \citep{oos00} existing between muscle contraction and ATP hydrolysis. The existence of such a loose coupling, in turn, assigns to the environmental thermal energy an essential role. In a previous paper~\citep{buo03}, it has been  shown that a model based on a washboard-type potential can be designed in a way that, by a suitable tuning of the involved parameters, one can fit the available data concerning the actomyosin interaction that is responsible for the muscle contraction mechanism (cf.~\cite{kit99,ish00,kit01} and the references therein).
\par
The mechanism underlying actomyosin mechanical interaction as a result of the energy supply
provided by ATP hydrolysis, and consequent transformation to ADP and Pi, has attracted a great interest not only within the realm of purely biological sciences but also among specialists of statistical mechanics and soft condensed matter. The variety of proposed theories~(\cite{mag93,pro94,ast97,mog98} and the references therein) share the aim of constructing models able to produce directional motion of a Brownian particle.
\par
In the present paper we construct a Brownian motor that embodies a fluctuating potential ratchet within an idealized cycle of  chemical reactions ensuing from ATP hydrolisis. The overall aim is to relate the net displacement of the myosin head to the amount of consumed energy in order to test whether the experimentally recorded displacements  are compatible with the available total energy. 
While postponing to future papers data fittings and numerical evaluations, here we focus on working out analytical expressions  for the duration of the ATP cycle, for the Gibbs free energy and for the net displacement of the myosin head. By exploiting the kinetic characterization of the heat bath and the energetics of thermal ratchet models along the lines indicated in  \cite{sek97}, a formula will also be obtained for the amount of energy consumed during the ATP cycle.
%
\section{The Model}
The aim of our model is to describe the dynamics of a myosin head during its interaction with
actin molecules. Here, the emphasis will be put on the construction of the model by
means of rigorously quantitative arguments that will lead us to rigorous equations and explicit
formulas. The implementation of the obtained mathematical results for data fitting and prediction
purposes will be the object of future endeavors. It should be pointed out that, differently from
previous papers (see~\cite{buo03,shipc}), our present approach focuses on energy
features of a chemically driven idealized gadget whose behavior is meant to mimic essential
features of myosin head's dynamics and of related energy properties.
\par
Since the myosin motion occurs in a liquid environment, the effect of drag viscous friction must be
included in the model. Moreover, on accounts of its size (see, for instance,~\cite{ishdm}), a multiplicity of forces due to thermal agitation affect the dynamics of the myosin head, that in the sequel we shall often refer to as to \lq\lq the particle\rq\rq. Such forces, all randomly acting on the microscopic scale, are generally synthetized in a unique resulting random force. Hence, a random component will be included by us in the equation modeling the dynamics of the particle.  Hereafter, we shall assume that the actomyosin interaction is of a conservative nature, and hence originating from a potential $U(x,t)$, that we are viewing a space-time dependent. Denoting by $x(t)$ the myosin head's position at time $t$ under the customary point-size approximation, by $m$ its mass and by $\rho$ the existing viscous friction coefficient, the equation of motion is assumed to be
\begin{equation}\label{Eq.moto}
m\ddot{x}=-\rho\dot{x}-\frac{\partial U(x,t)}{\partial x}+ \sqrt{2\rho k_BT}\,\xi(t)
\end{equation}
where $k_B$ is Bolzmann constant, $T$ is the absolute temperature and $k_BT$ (measured in pN $\cdot$ nm) is the thermal energy. The quantity $2\rho k_BT$ originates from the fluctuation-dissipation theorem that also implies that $\xi(t)$ is a zero-mean, delta-correlated Gaussian noise with unit intensity coefficient:
\begin{equation}\label{RumoreBianco}
\langle \xi (t)\rangle = 0; \qquad
\langle \xi (t)\xi (t')\rangle =\delta (t-t').
\end{equation}
\par
Eq.~(\ref{Eq.moto}) alone does not suffice to describe the dynamics of the myosin
head. Indeed, such dynamics is heavily biased by the chemical cycle elicited by ATP
hydrolization, one of its effects being changes of potential $U(x,t)$ in a fashion that is
independent of the position of the myosin head. For our modeling purposes, we assume that ATP
hydrolysis cycle is composed of three chemical states.  These are depicted in
Fig.~\ref{CicloChimico}, in which A, M, and Pi denote actin, myosin, and inorganic phosphate,
respectively.
\par
It is worth pointing out explicitly that, as it is often customary when formulating a model of a complex phenomenon, the 3-state model proposed here is a drastic simplification of the underlying biochemical and biophysical mechanisms. Nevertheless, such simplification catches the essential feature of the dynamics of the real phenomenon. Indeed, it depicts the random ``local'' fluctuations as well as the abrupt changes corresponding to detachment and attachment of the myosin head. What is even more relevant, is that via our model we obtain exact formulas, which is an important propedeutic tool aiming at quantitative data fitting of the considered phenomenon. Furthermore, it should not pass unnoticed that the presently proposed model involves some twelve parameters (six transition rates, extreme values, period and asymmetry parameter of the potential, viscous friction coefficient) so that it would be unrealistic at the present stage to increase further the complexity of the model if one wishes to make it suitable for fitting purposes.
\begin{figMacPc}
			{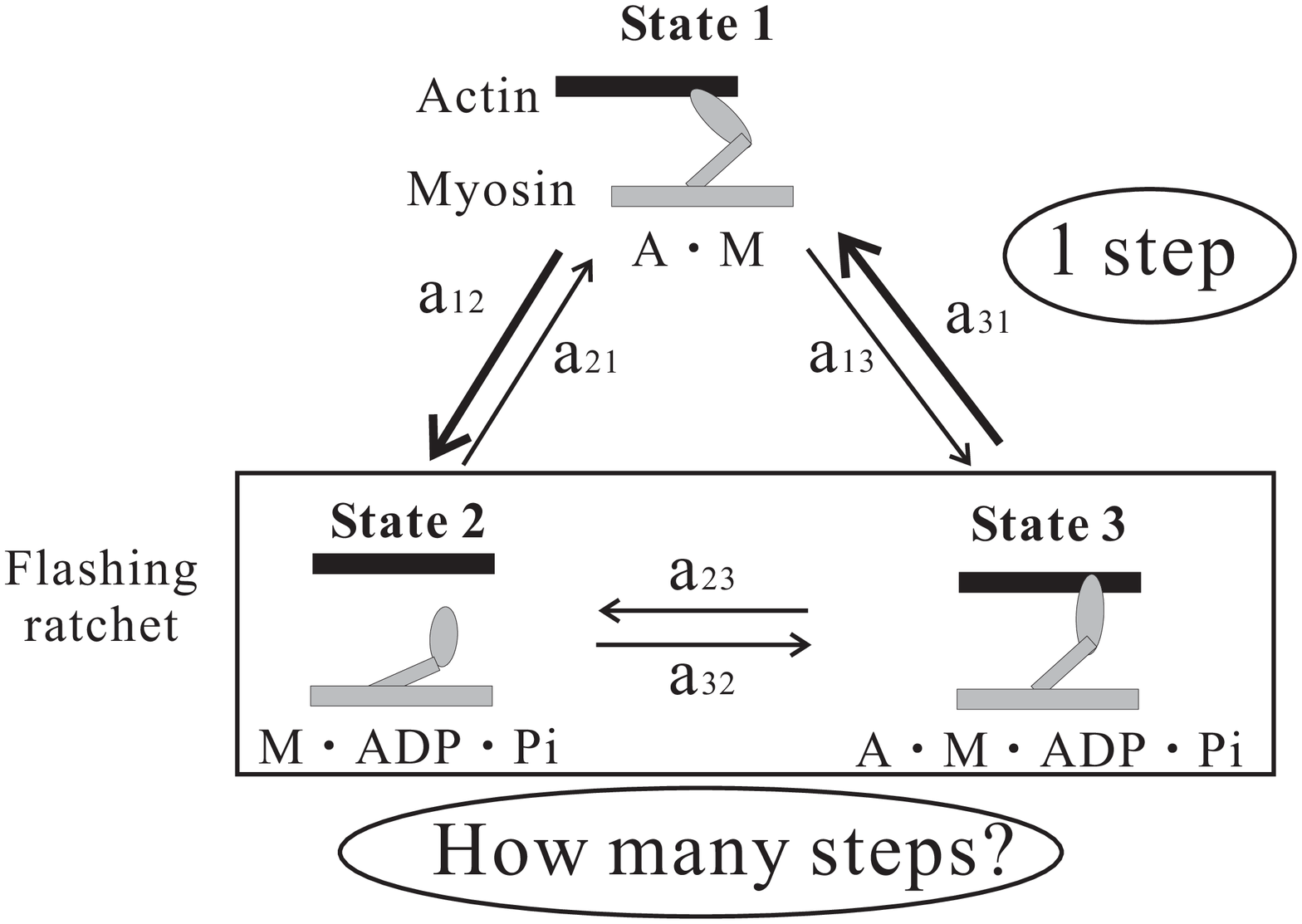}
			{7 cm}
			{The diagram of the chemical reaction route.}
			{CicloChimico}
\end{figMacPc}
\par
The transition of the chemical state from state \indice{i} to state \indice{j} occurs
according to a Poisson process with rates $a_{ij}$ for $i\neq j$. The relations among transition rates $a_{ij}$, equilibrium constants $K_{ij}$ and equilibrium velocities $k_{ij}$ are indicated in Table~\ref{Tabella1}.
\begin{tabMacPc}
			{Equilibrium constants, velocity constants, and transition rates of the chemical reaction route.}
			{
			\begin{center}
			\begin{tabular}{|c|l|l|}\hline
Equilibrium constant           & Velocity constant                   & Transition rate
($\mbox{s}^{-1}$) 
\\ \hline\hline
                       & $k_{12}$\numdim{}{M$^{-1}$s$^{-1}$} & $a_{12}=k_{12}$\conc{ATP} \\
$K_{12}=k_{12}/k_{21}$ &                                     &                           \\
                       & $k_{21}$\numdim{}{M$^{-1}$s$^{-1}$} & $a_{21}=k_{21}$\conc{A}   \\ 
\hline
                       & $k_{23}$\numdim{}{M$^{-1}$s$^{-1}$} & $a_{23}=k_{23}$\conc{A}   \\
$K_{23}=k_{23}/k_{32}$ &                                     &                           \\
                       & $k_{32}$\numdim{}{s$^{-1}$}         & $a_{32}=k_{32}$           \\
\hline
                       & $k_{31}$\numdim{}{s$^{-1}$}         & $a_{31}=k_{31}$           \\
$K_{31}=k_{31}/k_{13}$ &                                     &                           \\
                       & $k_{13}$\numdim{}{M$^{-2}$s$^{-1}$} &
$a_{13}=k_{13}$\conc{ADP}\conc{Pi}
\\ 	
\hline
			\end{tabular}
			\end{center}
			}
			{Tabella1}
\end{tabMacPc}
\par
We assume that when the chemical state at time $t$ is state \indice{i}, then the potential function
$U(x,t)$ is:
\begin{equation}\label{Potenziale}
U(x,t)=U_i(x)=\left\{
\begin{array}{ll}
\begin{displaystyle}-\frac{U_i}{\lambda _a}(x-\lambda _a)\end{displaystyle}
,&0\leq x[\mbox{mod }\lambda _b]<\lambda _a \\
\\
\begin{displaystyle}\frac{U_i}{\lambda _b-\lambda _a}(x-\lambda _a)\end{displaystyle}
,&\lambda _a\leq x[\mbox{mod }\lambda _b]<\lambda _b,
\end{array}
\right.
\end{equation}
where $\lambda _b$ is the length of an actin molecule and $\lambda _a$ is chosen
greater than $\lambda _b/2$ to break the symmetry of the potential function. We are thus invoking a periodic saw-tooth type potential to characterize each chemical state. When the particle moves from one potential well to a neighboring one, we conventionally say that the particle has moved one \lq\lq step\rq\rq. The net displacement of the particle at the end of an ATP cycle will thus be expressed as the difference between the number of steps moved in the positive direction and the number of steps moved in the negative direction, times the length $\lambda _b$ of one step. As sketched in Fig.~\ref{CicloChimico}, a relevant question relates to the number of steps during an ATP cycle. This figure also indicates that, at the end of the ATP cycle, one step in the prevailing direction  is performed by the myosin head, which is a consequence of the allosteric conformational change following the release of  Pi. Hereafter, $\lambda_a$, $\lambda _b$ and all other involved lengths will be expressed in nanometers. The value $U_i$ is the potential height for the chemical state \indice{i}. Since the binding strength 
between the myosin head and the actin filament is strong in states $i=1,3$ and weak in state $i=2$,
we choose the potential maxima as $U_1\ge U_3>U_2$. 
\par
We now remark that it is conceivable to assume that the motion of the particle can be described under an overdamped approximation. Indeed, the viscous friction coefficient is given by $\rho=$ \numdim{90}{pN $\cdot$ ns/nm}, the mass of the myosin head is $m=$ \numdim{2.2 \cdot 10^{-22}}{Kg} and $\lambda_b=$ \numdim{6}{nm} (see, for instance,~\cite{buo03,ishdm,ish00}, respectively). Hence, denoting by $M=$ \numdim{100}{pN $\cdot$ nm} a sure upper bound to the potentials $U_1,\,U_2,\,U_3$, one has 
$$
\frac{m}{\rho^2\lambda_b^2}M=\frac{2.2\cdot 10^{-1}\cdot 100}{(9\cdot 10\cdot 6)^2}\simeq 10^{-4}.
$$
The non-dimensional quantity on the left-hand side being much smaller than 1, justifies the above-mentioned  approximation~\citep{rei00}. Since the motion is overdamped, the
inertial component in Eq.~(\ref{Eq.moto}) can be disregarded, so that the dynamics of the particle 
will henceforth be modeled by the following Langevin equation:
\begin{equation}\label{Eq.Langevin}
\dot{x}=-\frac{1}{\rho}\frac{\partial U(x,t)}{\partial x}+ \sqrt{2D}\xi (t)
\end{equation}
where $D=k_BT/\rho$ is Einstein's diffusion coefficient.
\par
Let now $Z(t)=\left[X(t),I\right]$ be a random vector whose components $X(t)$ and $I$ have
the following meaning: $X(t)$ denotes the position of the myosin head at time $t$ and $I$ is
a random variable, defined over the set $\left\{1,2,3\right\}$, denoting the chemical state of
the actomyosin system. Then,
\begin{displaymath}
f_i(x,t) := \frac{\partial}{\partial x}P\left\{X(t)<x;I=i\right\} \qquad\mbox{(\indice{i})} 
\end{displaymath}
is the joint probability density function (pdf) of $\left\{X(t)\in(x,x+dx)\cap I=i\right\}$. Differently stated, $f_i(x,t)dx$ is
the probability that the myosin head at time $t$ is near $x$ and that at the same time the
actomyosin system is in chemical state $i$. The corresponding Fokker-Planck equation
for $f_i(x,t)$ for the chemical state \indice{i} is
\begin{equation}\label{FokkerPlank}
\frac{\partial f_i}{\partial t}(x,t)=
-\left(\sum^3_{\stackrel{j\neq i}{j=1}}a_{ij}\right) f_i(x,t) 
+\sum^3_{\stackrel{k\neq i}{k=1}}a_{ki} f_k(x,t)
+\frac{\partial}{\partial x}\left[\frac{1}{\rho} \frac{dU_i}{dx}(x) f_i(x,t)\right]
+ D\frac{\partial ^2f_i}{\partial x^2} (x,t).
\end{equation}
For \indice{i},
\begin{displaymath}
p_i(x,t)=\sum_{r=-\infty}^{+\infty}f_i(x+r\lambda_b,t)
\end{displaymath}
is a periodic function of period $\lambda_b$ that is a solution of
Eq.~(\ref{FokkerPlank}) in which the space variable is made to vary in the interval  $(0,\lambda_b)$. Under such reduced
dynamics, the existence of a non trivial stationary regime is suggested by  physical
considerations~\citep{rei00}. In this stationary regime, $p^{st}_i(x)\equiv p_i(x)$ for states \indice{i} are the
joint pdf's of the position $X$ of the myosin head and of the states $I$ of the chemical system:
\begin{equation}\label{pdfStazionarie}
p_i(x) = \frac{d}{d x}P\left\{X<x;I=i\right\}\qquad\mbox{(\indice{i})}. 
\end{equation}
Such functions satisfy
\begin{displaymath}
0=-\left(\sum^3_{\stackrel{j\neq i}{j=1}}a_{ij}\right) p_i(x) 
+\sum^3_{\stackrel{k\neq i}{k=1}}a_{ki} p_k(x)
+\frac{d}{dx}\left[\frac{1}{\rho} U^\prime_i (x)p_i(x)\right]
+ Dp^{\prime\prime}_i(x)
\qquad\mbox{(\indice{i})}
\end{displaymath}
or
\begin{equation}\label{eq_pdfStazionarie}
0=-\left(\sum^3_{\stackrel{j\neq i}{j=1}}a_{ij}\right) p_i(x) 
+\sum^3_{\stackrel{k\neq i}{k=1}}a_{ki} p_k(x)
-S^\prime_i(x)
\qquad\mbox{(\indice{i})},
\end{equation}
where
\begin{equation}\label{Flussi}
S_i(x)=-\left[\frac{1}{\rho}U^\prime_i(x)+D\frac{d}{dx}\right]p_i(x)
\qquad\mbox{(\indice{i})}
\end{equation}
denote the probability currents.
\par
Note that adding term by term Eqs.~(\ref{eq_pdfStazionarie}) one obtains: 
\begin{displaymath}
S^\prime_1(x)+S^\prime_2(x)+S^\prime_3(x)=0,
\end{displaymath}
where prime means space derivative. Hence,
\begin{equation}\label{FlussoTotale}
S(x):=S_1(x)+S_2(x)+S_3(x)=S,
\end{equation}
with $S$ a constant denoting the total probability current of the myosin head at each
$x\in(0,\lambda_b)$. Instead,
\begin{equation}\label{ProbabilitaTotale}
p(x):=p_1(x)+p_2(x)+p_3(x)
\end{equation}
is the pdf of the position of myosin head in $(0,\lambda_b)$. 
\par
We conclude this section by working out some results use of which will be made in the sequel.
\begin{proposizione}\label{IntgPrimo}
Let $g(x)$ be a periodic continuous function of period $\lambda_b$ differentiable in $(0,\lambda_b)$
apart, at most, at an inner point $\lambda_a$ where, however, there exist bounded left and right
derivatives. Then,
\begin{displaymath}
\int_0^{\lambda_b}g^\prime(x)\,dx=0.
\end{displaymath}
\end{proposizione}
\begin{dimos}
Due to the assumed continuity and periodicity of $g(x)$ one has
$g(0^+)=g(0)=g(\lambda_b)=g(\lambda_b^-)$ and $g(\lambda_a^-)=g(\lambda_a^+).$
Hence, 
\begin{eqnarray*}
\int_0^{\lambda_b}g^\prime(x)\,dx
&=&\int_0^{\lambda_a}g^\prime(x)\,dx+\int_{\lambda_a}^{\lambda_b}g^\prime(x)\,dx
=\left[g(\lambda_a^-)-g(0^+)\right]+\left[g(\lambda_b^-)-g(\lambda_a^+)\right]\\
&=&\left[g(\lambda_a^-)-g(\lambda_a^+)\right]+\left[g(\lambda_b^-)-g(0^+)\right]
=0.
\end{eqnarray*}
\end{dimos}
\begin{corollario}
For \indice{i} one has: 
\begin{eqnarray}\label{IntpSPrimo}
\int_0^{\lambda_b}p_i^\prime(x)\,dx&=&0 \label{intpiprimo}\\ 
\int_0^{\lambda_b}S_i^\prime(x)\,dx&=&0.\label{intSprimo}
\end{eqnarray}
\end{corollario}
\begin{dimos}
The proof follows immediately by nothing that the probability density functions~(\ref{pdfStazionarie}) satisfy the assumptions of Proposition~\ref{IntgPrimo}. Indeed, they are expressed as linear combinations of exponential functions in which the combination coefficients follow by imposing that the Fokker-Plank equations are satisfied and that continuity conditions hold at the end points $0$ and $\lambda_b$ and at the potential minimum $\lambda_a$. Similar conditions also hold for the probability currents. (See, for instance, \cite{ast94} for a detailed argument). 
\end{dimos}

In conclusion, we point out that from Eq.~(\ref{ProbabilitaTotale}) and from Eq.~(\ref{intpiprimo}), that holds for \indice{i}, one obtains:
\begin{equation}\label{intpprimo}
\int_0^{\lambda_b}p^\prime(x)\,dx=0
\end{equation}

%
\section{State occupation probabilities and  Gibbs Free energy}
The probability of the state \indice{i} in the stationary regime is given by:
\begin{equation}
P_i=\int _0^{\lambda _b}p_i(x)dx
\end{equation}
which is independent of both $x$ and $t$. Then, we have
\begin{proposizione}
The stationary state probabilities satisfy the following algebraic linear system:
\begin{equation}\label{Pi}
\left\{
\begin{array}{rrrr}
0=&-(a_{12}+a_{13})P_1 &+\ a_{21}P_2          &+\ a_{31}P_3 \\
0=&a_{12}P_1           &-\ (a_{21}+a_{23})P_2 &+\ a_{32}P_3\\
0=&a_{13}P_1           &+\ a_{23}P_2          &-\ (a_{31}+a_{32})P_3 
\end{array}
\right.
\end{equation}
\end{proposizione}
\begin{dimos}
System~(\ref{Pi}) immediately follows by integration of Eq.~(\ref{eq_pdfStazionarie}) for 
\indice{i} with respect to $x$ from $0$ to $\lambda _b$ and by recalling Eq.~(\ref{intSprimo}). 
\end{dimos}
\par
An alternative and more straightforward proof, that will also allow us to introduce some necessary notation and a useful result, will now be given.
\begin{dimos}
Let $a=(a_{12}+a_{21}+a_{23}+a_{32}+a_{31}+a_{13})^{-1}$ and let $\Delta$ be a positive quantity,
with dimension of time such that $\Delta\ll a$. Moreover, let $P_{ij}$ be the probabilities that
the $3$-state model undergoes  a transition from state $i$ to state $j$ 
(\indici{i}{j}) during $\Delta$. The probability for the system to be in state \indice{i} at time $t$ satisfies the following equation: 
\begin{equation}\label{Pit}
P_i(t+\Delta)=P_1(t)P_{1i}+P_2(t)P_{2i}+P_3(t)P_{3i}.
\end{equation}
In stationary condition, Eq.~(\ref{Pit}) yields:
\begin{displaymath}
P_i=P_1P_{1i}+P_2P_{2i}+P_3P_{3i}=P_iP_{ii}+\sum^3_{\stackrel{j\neq
i}{j=1}}P_jP_{ji}
\end{displaymath}
or:
\begin{displaymath}
0=(P_{ii}-1)P_i+\sum^3_{\stackrel{j\neq i}{j=1}}P_jP_{ji}.
\end{displaymath}
From the normalization conditions $P_{i1}+P_{i2}+P_{i3}=1$, for \indice{i}, and from the remark
that  for $i\neq j$ one has 
\begin{equation}\label{Pij}
P_{ij}=\Delta\cdot a_{ij},
\end{equation}
system~(\ref{Pi}) finally follows.
\end{dimos}
The homogeneous linear system~(\ref{Pi}) has rank $2$, so that it admits an infinity number of
solutions. The states occupation probabilities are obtained via the normalization condition
$P_1+P_2+P_3=1$. We further note that system ~(\ref{Pi}) admits the invariant 
\begin{equation}\label{J}
J_C=a_{12}P_1-a_{21}P_2=a_{23}P_2-a_{32}P_3=a_{31}P_3-a_{13}P_1, 
\end{equation}
representing the frequency of the ATP cycle in   the $3$-state model under the stationary regime. Its reciprocal
$T_C=1/J_C$ expresses the averaged time period of one ATP cycle. By multiplying both sides
of Eq.~(\ref{J}) by $\Delta$, one obtains another invariant expressed in terms of occupation and of state transition probabilities during time interval $\Delta$: 
\begin{equation}\label{Jdelta}
J_C\Delta=P_{12}P_1-P_{21}P_2=P_{23}P_2-P_{32}P_3=P_{31}P_3-P_{13}P_1. 
\end{equation}
\par
We are now in the position to calculate Gibbs free energies $\Delta G_{12}$, $\Delta G_{23}$,
$\Delta G_{31}$ of all chemical reactions involved in the $3$-state model. Indeed, assuming that
the concentrations of the various reactants are proportional to the occupation
probabilities of their respective states, one obtains: 
\begin{eqnarray}
\Delta G_{12}&=& -k_BT\ln K_{12} + k_BT\ln \frac{[\mbox{M}\cdot \mbox{ADP}\cdot\mbox{Pi}]
[\mbox{A}]}{[\mbox{A}\cdot\mbox{M}] [\mbox{ATP}]}  =k_BT\ln \frac{P_2[\mbox{A}]}{K_{12}P_1
[\mbox{ATP}]}\nonumber \\ 
&=&k_BT\ln \frac{a_{21}P_2}{a_{12}P_1}
\label{eq:G_12} \\
\nonumber \\
\Delta G_{23}&=&-k_BT\ln K_{23}+k_BT\ln \frac{[\mbox{A}\cdot\mbox{M}\cdot
\mbox{ADP}\cdot\mbox{Pi}]}{[\mbox{M}\cdot\mbox{ADP}\cdot\mbox{Pi}][\mbox{A}]}  =k_BT\ln
\frac{P_3}{K_{23}P_2[\mbox{A}]}\nonumber\\
&=&k_BT\ln \frac{a_{32}P_3}{a_{23}P_2}
\label{eq:G_23} \\
\nonumber \\
\Delta G_{31}&=&-k_BT\ln K_{31}+k_BT\ln
\frac{[\mbox{A}\cdot\mbox{M}][\mbox{ADP}][\mbox{Pi}]}{[\mbox{A}\cdot\mbox{M}\cdot{ADP}\cdot
\mbox{Pi}]}  =k_BT\ln \frac{P_1[\mbox{ADP}][\mbox{Pi}]}{P_3K_{31}}\nonumber\\
&=&k_BT\ln \frac{a_{13}P_1}{a_{31}P_3}
\label{eq:G_31}
\end{eqnarray}
where use of the relations indicated in Table~(\ref{Tabella1}) has been made. 
\par
Formulae ~(\ref{eq:G_12}),~(\ref{eq:G_23}) and~(\ref{eq:G_31}) explicitly show the dependence of
Gibbs free energies on transition rates as well as on states occupation probabilities in the stationary regime.
\begin{proposizione}
Gibb's total free energy $\Delta G$ of an ATP cycle depends only on the transition rates of
the $3$-states model. 
\end{proposizione}
\begin{dimos}
Adding sides by sides Eqs.~(\ref{eq:G_12})~--~(\ref{eq:G_31}) and recalling the relations
in Table~\ref{Tabella1}, one obtains 
\begin{equation}
\Delta G=\Delta G_{12}+\Delta G_{23}+\Delta G_{31}=
k_BT\ln \frac{[\mbox{ADP}][\mbox{Pi}]}{K_{12} K_{23} K_{31} [\mbox{ATP}]}
=k_BT\ln\frac{a_{21}a_{32}a_{13}}{a_{12}a_{23}a_{31}}.
\end{equation}
\end{dimos}
%
\section{Step numbers and energy consumption}
The following proposition extends  to the
case of potential~(\ref{Potenziale}) the well-known property (see, for instance, \cite{ris84}) that the velocity and probability current are proportional. 
\begin{proposizione}\label{velocita}
The average speed of the myosin head in the $3$-state model is given by
\begin{equation}
\overline{v}:= E\left[\dot{x}(t)\right]=\lambda_b S
\end{equation}
where $S$ is the probability current defined in Eq.~(\ref{FlussoTotale}).
\end{proposizione}
\begin{dimos}
Taking expectation on both sides of Eq.~(\ref{Eq.Langevin}) and recalling the first
of~(\ref{RumoreBianco}) we obtain:
\begin{displaymath}
\overline{v}=-\frac{1}{\rho}E\left[\frac{\partial U(x,t)}{\partial x}\right].
\end{displaymath}
The expectation on the right-hand-side can be calculated by decomposition in terms of the three
mutually exclusive and exhaustive chemical states. Hence: 
\begin{equation}\label{velocita2}
\overline{v} = -\frac{1}{\rho}\sum_{i=1}^3P_iE\left[\left.\frac{\partial U(x,t)}{\partial x}\right| I = i\right]    = -\frac{1}{\rho}\sum_{i=1}^3P_i\int_0^{\lambda_b} U_i^\prime(x)\frac{p_i(x)}{P_i}\, dx 
= -\frac{1}{\rho}\sum_{i=1}^3\int_0^{\lambda_b} U_i^\prime(x)p_i(x)\, dx\,.
\end{equation}
Recalling the definition~(\ref{Flussi}) of the probability current when the system is in state
\indice{i}, and making use of~(\ref{FlussoTotale}) and~(\ref{ProbabilitaTotale}),
from~(\ref{velocita2}) we obtain: 
\begin{eqnarray*}
\overline{v} 
&=& \sum_{i=1}^3\int_0^{\lambda_b}\left[S_i(x) + Dp^\prime_i(x)\right]\, dx
  = \int_0^{\lambda_b}\sum_{i=1}^3 S_i(x)\, dx + D\int_0^{\lambda_b}\sum_{i=1}^3p^\prime_i(x)\, dx
  \\
&=& \int_0^{\lambda_b}S(x)\, dx + D\int_0^{\lambda_b}p^\prime(x)\, dx
\,.
\end{eqnarray*}
Use of Eqs.~(\ref{FlussoTotale}) and~(\ref{intpprimo}) complete the proof. 
\end{dimos}
From Proposition~\ref{velocita} we obtain the following
\begin{corollario}
The net average number $\overline{n}$ of steps performed by the myosin head is given by
\begin{equation}\label{step}
\overline{n} = \frac{S}{J_C},
\end{equation}
where $S$ is the total probability current of the myosin head and $J_C$ is the frequency of the ATP cycle in the $3$-state model.
\end{corollario}
\begin{dimos}
The net displacement of the myosin head during one ATP cycle of duration $T_C$ is
\begin{displaymath}
\overline{v}\cdot T_C = \lambda_b S\cdot T_C = \lambda_b S\cdot \frac{1}{J_C}.
\end{displaymath}
Dividing by the length $\lambda_b$ of a step we are led to~(\ref{step}).  
\end{dimos}
\par
Next, we shall calculate the mean energy consumed during one ATP cycle for the $3$-state model, which depends on the potential differences among the states. When the particle occupies position $x$ at the time $t$ when a transition takes place from state $i$ to state $j$, the energy consumed is (see~\cite{sek97})  
\begin{equation}
H(x,t)\equiv\Delta U_{ij}(x)=U_j(x)-U_i(x).
\end{equation}
Let now $\tau \geq 0$ denote an arbitrary instant. In $(\tau,\tau+\Delta)$ a state transition certainly occurs. The energy consumed during such transition is a function of (random) position $X\in(o,\lambda_b)$ of the particle at time $\tau$, of the (random) occupied state $I\in\{1,2,3\}$ in the $3$-state model and on the (random) direction of the transition $I\to J$, where $J\in\{1,2,3\}$: 
\begin{equation}\label{EnergiaImmessa}
H(X,\tau)=H_{IJ}(X)=\Delta U_{IJ}(X).
\end{equation}
\begin{proposizione}
The average energy $\overline{H}$ during a transition occurring in $(\tau,\tau+\Delta)$ is given by:
\begin{eqnarray}\label{Hsegnato}
\overline{H}:= E\left[H(X,\tau)\right]
&=&P_1P_{12}(H_{12}+H_{21})+P_2P_{23}(H_{23}+H_{32})+P_3P_{31}(H_{31}+H_{13})+\nonumber \\
& &\qquad -J_C\Delta(H_{13}+H_{32}+H_{21})
\end{eqnarray}
where
\begin{equation}
H_{ij}=E\left[H_{ij}(X)\right]=\int_0^{\lambda_b}H_{ij}(x)\frac{p_i(x)}{P_i}\, dx
=\int_0^{\lambda_b}\Delta U_{ij}(x)\frac{p_i(x)}{P_i}\, dx\qquad \mbox{\emph{(}\indici{i}{j}\emph{)}},
\end{equation}
the $P_i$'s are the stationary state probabilities and $P_{ij}$'s are the probabilities of the transition $i\to j$ in time interval $\Delta$.
\end{proposizione}
\begin{dimos}
A repeated use of averages of conditional expectations yields:
\begin{eqnarray*}
E\left[H(X,\tau)\right]
&=&\sum_{i=1}^{3}P_i\,E\left[H(X,\tau)\mid I=i\right]
=\sum_{i=1}^{3}P_i\,E\left[H_{iJ}(X)\right] \\ 
&=&\sum_{i=1}^{3}P_i\sum_{j=1}^{3}P_{ij}E\left[H_{iJ}(X)\mid i \to j \right]
=\sum_{i=1}^{3}P_i\sum_{j=1}^{3}P_{ij}E\left[H_{ij}(X)\right]
\\
&=&\sum_{i=1}^{3}\sum_{j=1}^{3}P_iP_{ij}H_{ij}.
\end{eqnarray*}
To complete the proof we notice that $H_{ii}=0$ for \indice{i} and that, by virtue of the invariant~(\ref{Jdelta}), one has
\begin{displaymath}
P_2P_{21}=P_1P_{12}-J_C\Delta;\qquad P_3P_{32}=P_2P_{23}-J_C\Delta;
\qquad P_1P_{13}=P_3P_{31}-J_C\Delta.
\end{displaymath}
\end{dimos}
We are now in the position to calculate the mean energy consumed during one ATP cycle. 
\begin{proposizione}
The total mean energy $\mathcal{H}$ consumed during one ATP cycle is 
\begin{eqnarray}\label{H}
\mathcal{H}
&=&\frac{1}{J_C}\left[P_1a_{12}(H_{12}+H_{21})+P_2a_{23}(H_{23}+H_{32})+P_3a_{31}(H_{31}+H_{13})\right]+\nonumber
\\  
& &\qquad -(H_{13}+H_{32}+H_{21}),
\end{eqnarray}
where $a_{ij}$ (\indici{i}{j}; $i\neq j\,$) is the transition rate from state $i$ to state $j$.
\end{proposizione}
\begin{dimos}
The total consumed energy is expressed as a random sum of random variables, the number of terms to be added being the random number $N$ of state transition that occur in the $3$-state model during one ATP cycle. Furthermore, each such term is a random variable having the same distribution function as $H(X,\tau)$. Hence, the mean total energy $\mathcal{H}$ is obtained as the product of the mean value $\overline{H}$ of the random variable $H(X,\tau)$ times the average number $\overline{N}$ of transition during one ATP cycle. Since the mean duration of one ATP cycle is $T_C=1/J_C$, we obtain:
\begin{displaymath}
\overline{N}:=E\left[N\right]=\sum_{i=1}^{3}P_iE\left[N\mid I=i\right]
\sum_{i=1}^{3}P_i\sum_{j=1}^{3}P_{ij}\frac{T_C}{\Delta}=
\frac{T_C}{\Delta}\sum_{i=1}^{3}P_i\sum_{j=1}^{3}P_{ij}=
\frac{T_C}{\Delta}\sum_{i=1}^{3}P_i=
\frac{T_C}{\Delta}=\frac{1}{J_C\Delta}.
\end{displaymath}
Use of~(\ref{Pij}) and~(\ref{Hsegnato}) completes the proof.
\end{dimos}
\section{Concluding remarks}
The necessity of relating the dynamics of the displacements of Myosin II heads during its interaction with actin filaments ---as observed in the experimental work by T. Yanagida and his group--- with the energy available during an ATP cycle has been the object of the present paper. To this purpose, we have designed a Brownian motor based on a fluctuating potential ratchet working as a gadget driven by a cycle of chemical reactions among hypothesized three different states of the actomyosin system. We have obtained rigorous analytical expressions for the duration of the ATP cycle, for the Gibbs free energies, for the net displacement of the myosin head and for the average energy consumed during an ATP cycle. Use of these theoretical results will be made in future endeavors in order to test in a quantitative fashion whether the existence of multiple steps is compatible with energy balance requirements. 
%

%

\begin{thebibliography}{99}
%
\bibitem[Sekimoto(1997)]{sek97} 
      Sekimoto K., 1997.  
		Kinetic Characterization of Heat Bath and the Energetics of Thermal Ratchet Models.  
      J. Phys. Soc. Jpn. 66, 1234-1237.
%
\bibitem[Oosawa(2000)]{oos00}
	Oosawa F., 2000.
	The loose coupling mechanism in molecular machines of living cells. 
	Genes to Cells 5, 9-16.
%
\bibitem[Buonocore and Ricciardi(2003)]{buo03}
	Buonocore A., Ricciardi L.M., 2003.
	Exploiting Thermal Noise for an Efficient Actomyosin Sliding Mechanism.
	Math. Biosci 182, 135-149.
%
\bibitem[Kitamura et al.(1999)]{kit99}
	Kitamura K., Okunaga M., Iwane A.H. Yanagida T., 1999.
	A single myosin head moves along an actin filament with regular steps of 5.3 
	nanometres.
	Nature 397, 129-134.
%
\bibitem[Ishii and Yanagida(2000)]{ish00}
	Ishii Y., Yanagida T., 2000.
	Single Molecule Detection in Life Science. 
	Single Mol. 1, 5-16.
%
\bibitem[Kitamura et al.(2001)]{kit01}
	Kitamura K., Ishijima A., Tokunaga M., Yanagida T., 2001.
	Single-Molecule Nanobiotechnology. 
	JSAP International 4, 4-9.
%
\bibitem[Magnasco(1993)]{mag93}
	Magnasco M.O., 1993.
	Forced Thermal Ratchets. 
	Phys. Rev. Lett. 71, 1477-1481.
%
\bibitem[Prost et al.(1994)]{pro94}
	Prost J., Chauwin J.F., Peliti L., Ajdari A., 1994.
	Asymmetric Pumping of Particles. 
	Phys. Rev. Lett. 72, 2652-2655.
%
\bibitem[Astumian(1997)]{ast97}
	Astumian R.D., 1997.
	Thermodynamics and Kinetics of a Brownian Motor.
	Science 276, 917-922.
%
\bibitem[Mogilner et al.(1998)]{mog98}
	Mogilner A., Mangel M., Baskin R.J., 1998.
	Motion of molecular motor ratcheted by internal fluctuations 	and protein friction. 
	Phys. Lett. A 237, 297-306.
%
\bibitem[Shimokawa and Sato(2001)]{shipc}
  Shimokawa T., Sato S., 2001.
  The analysis of a stochastic ratchet model.
  Technical Report of IEICE (in Japanese) 100, 63-70.
%
\bibitem[Ishiwatari(1997)]{ishdm}
	Ishiwatari S. (Ed.), 1997.
	Mechanism of Biological Molecular Motors. 
	Kyoritsu Shuppan, Tokyo (In Japanese).
%
\bibitem[Reimann(2002)]{rei00}
	Reimann P., 2002.
	Brownian motors: noisy transport far from equilibrium. 
	Phys. Rep. 361, 57-265.
%
\bibitem[Astumian and Bier(1994)]{ast94} 
      Astumian R.D., Bier M., 1994.
      Fluctuation Driven Ratchets: Molecular Motors.
      Phys. Rev. Lett. 72, 1766-1769.
%
\bibitem[Risken(1984)]{ris84}
	Risken H., 1984.
	The Fokker-Plank Equation. 
	Springer-Verlag, Berlin.
%
\end{thebibliography}
\end{document}